\documentclass[sigplan,10pt,nonacm]{acmart}

\usepackage[textsize=tiny]{todonotes}
\usepackage{algorithm2e}
\usepackage{indentfirst}
\usepackage{booktabs, makecell}
\usepackage[skins,breakable]{tcolorbox}
\AtBeginDocument{%
  }

\setcopyright{acmlicensed}
\copyrightyear{2026}
\acmYear{2026}
\acmDOI{XXXXXXX.XXXXXXX}
\acmISBN{978-1-4503-XXXX-X/2018/06}

\begin{document}

\title{Profiling the Energy Consumption of Serverless Functions with Joule Profiler}

\author{Jérémy Woirhaye}
\affiliation{%
  \institution{Inria / Univ.\,Lille / CNRS, CRIStAL}
  \city{Lille}
  \country{France}
}
\email{jeremy.woirhaye@inria.fr}

\author{François Gibier}
\affiliation{%
  \institution{Inria / Univ.\,Lille / CNRS, CRIStAL}
  \city{Lille}
  \country{France}
}
\email{francois.gibier@inria.fr}

\author{Anderson Andrei Da Silva}
\affiliation{%
  \institution{Inria / Univ.\,Lille / CNRS, CRIStAL}
  \city{Lille}
  \country{France}
}
\email{anderson-andrei.da-silva@inria.fr}

\author{Francescomaria Faticanti}
\affiliation{%
  \institution{Inria / Univ.\,Lille / CNRS, CRIStAL}
  \city{Lille}
  \country{France}
}
\email{francescomaria.faticanti@inria.fr}

\author{Thomas Ledoux}
\affiliation{%
  \institution{IMT Atlantique / Inria, LS2N}
  \city{Nantes}
  \country{France}
}
\email{thomas.ledoux@imt-atlantique.fr}

\author{Romain Rouvoy}
\affiliation{%
  \institution{Univ.\,Lille / Inria / CNRS, CRIStAL}
  \city{Lille}
  \country{France}
}
\email{romain.rouvoy@inria.fr}

\renewcommand{\shortauthors}{Woirhaye~\emph{et~al.}}

\begin{abstract}
Cloud providers and customers have widely adopted serverless computing as a convenient paradigm for deploying and executing functions on demand.
To do so, serverless platforms require provisioning an appropriate execution environment before a single line of the function's code runs. These environments consist of several layers, such as container engines, hypervisors, unikernels, and programming language runtimes.
While the literature has investigated the performance of these serverless platforms, it treats functions as black boxes, and the community lacks key insights into the environmental impacts of packaging applications as serverless functions.

This paper therefore empirically studies the energy efficiency of serverless functions deployable on serverless platforms.
We design an experimental benchmarking environment that lets stakeholders explore the impacts of the various layers involved in executing serverless functions.
We use it to evaluate 1,401 configurations, combining 9 execution environments, 7 language-runtime configurations, 11 workloads, and 3 input sizes, to answer three research questions: 
Are the most popular programming languages for serverless functions the most energy-efficient? 
What factors most affect their energy efficiency? 
What are the most energy-efficient configurations to deploy them? 
Our results show that one should first choose the programming language, then the language runtime, and only then the execution environment, which matters only for short-lived functions and whose best choice depends on the runtime.
Our benchmarking environment, experimental artifacts, raw measurements, and analysis code are publicly available.
\end{abstract}



\received{24 September 2026}
\received[revised]{12 March 2009}
\received[accepted]{5 June 2009}

\maketitle

\section{Introduction}\label{sec:intro}
Serverless computing, and more specifically the \emph{Function-as-a-Service} (FaaS) model, has become a pillar of modern cloud architectures.
It lets developers deploy small pieces of code, so-called functions, without managing the underlying infrastructure, which brings agility and can reduce operational costs.
But this convenience may come at the cost of reduced energy efficiency in cloud infrastructures because these functions run briefly, potentially wasting computing resources on management overhead.

Previous studies have measured and attributed the energy consumption and carbon footprint of serverless platforms~\cite{dasilvaAssessingPowerUsage2025, faasmeter, royHiddenCarbonFootprint2024, sharmaAccountableCarbonFootprints2024}, or reduced it through platform-level mechanisms such as runtime redesign~\cite{stojkovicEcoFaaSRethinkingDesign2024}, resource multiplexing~\cite{stojkovicMXFaaSResourceSharing2023}, energy- or carbon-aware scheduling~\cite{dasilva-acmsac25, rastegarEneXEnergyAwareExecution2024, jiang2024ecolife, chadha2023greencourier, aslanpour2022energy}, and energy-efficient hardware~\cite{byrne2022microfaas}. 
In all these works, however, they treat function implementation as a black box. 
This paper proposes a novel perspective on serverless energy efficiency by considering the impacts of implementation technologies, such as programming languages, runtime environments, and different virtualization alternatives (container engines, unikernels, and hypervisors).
More specifically, this paper addresses three research questions:
\textbf{RQ1.} Are the most popular programming languages for implementing serverless functions the most energy-efficient? 
\textbf{RQ2}. What factors most affect the energy efficiency of serverless functions?
\textbf{RQ3}. What are the most energy-efficient configurations to adopt when using serverless functions?


We answer these questions by designing and executing a middleware benchmark infrastructure that explores various dimensions of serverless environments and provides empirical evidence on the effects of different deployment factors, such as programming languages and runtime environments. 
In particular, we evaluate 1,401 configurations (60,780 runs) that combine different execution environments, language-runtime configurations, workloads, and input sizes. 
%
%
%
We believe this work provides stakeholders with useful insights into the trade-offs involved in selecting deployment configurations that reduce the energy consumption of their services and infrastructure.
Prior work has typically either measured energy at the platform level without isolating execution phases~\cite{dasilvaAssessingPowerUsage2025, royHiddenCarbonFootprint2024, sharmaAccountableCarbonFootprints2024}, or compared environments on latency alone without energy~\cite{dinhtuanUnikernelsVsContainers2025, goethalsFunctionalPerformanceBenchmark2022, moebiusAreUnikernelsReady2024}. 
To the best of our knowledge, this work is the first to provide a phase-level comparison of both energy efficiency and performance across the full execution stack, from bare metal to unikernels, while also considering seven language–runtime configurations. 
%
The key contributions of this paper are:

\begin{itemize}
    \item A systematic, phase-level evaluation of energy consumption across combinations of execution environments, language-runtime configurations, and workloads.
    \item A characterization of how execution-environment overhead affects energy consumption beyond bare-metal execution: differences between environments are dominated by boot and teardown rather than computation, with teardown alone accounting for 40--47\% of lifecycle energy for container-based environments and one third for Nanos running on Firecracker.
    \item Evidence that the choice of language runtime has a larger impact on energy consumption than the choice of container engine or hypervisor.
    \item The finding that no single execution configuration is optimal for all workloads: the most energy-efficient environment depends on the hosted runtime, and the time required to amortize environment overhead ranges from 25\,ms to nearly 900\,ms.
\end{itemize}


\section{Background \& related works}\label{sec:back}
The development of cloud computing and serverless technologies has diversified execution environments, each offering different trade-offs among performance, isolation, and efficient resource utilization. 
These execution environments rely on different virtualization mechanisms that play an important role in serverless computing.
However, multiplying abstraction layers (containerization, hardware virtualization, etc.) raises questions about their actual impact on performance and energy consumption.
Prior work suggests two expectations: a larger performance effect when changing the programming language, of about an order of magnitude on bare metal~\cite{pereiraEnergyEfficiencyProgramming2017}, and a smaller performance effect when changing the virtualization mechanism, concentrated in the cold start~\cite{goethalsFunctionalPerformanceBenchmark2022, moebiusAreUnikernelsReady2024, dinhtuanUnikernelsVsContainers2025}. However, neither effect has been quantified in energy across the full serverless lifecycle, and no prior work isolates the effect of the language runtime for the same source code. This paper addresses these two gaps.
In the rest of this section, we define key concepts and discuss related work on these topics.

\subsection{Virtualization technologies}
\textbf{Containers} are one of the most common ways to isolate serverless functions. Open-source platforms such as Knative~\cite{knative}, OpenFaaS~\cite{openfaas}, and Apache OpenWhisk~\cite{openwhisk} package each function as a container image started when an event arrives~\cite{8817155}, and commercial platforms such as AWS Lambda, Azure Functions, and Google Cloud Run functions also accept container images~\cite{awslambda, azurefunctions, gcprunfunctions}. Containers do not virtualize the hardware: applications run directly on the host kernel, isolated by namespaces and limited by cgroups. Sharing the kernel, however, makes their isolation weaker than hardware virtualization.

\textbf{Virtual machines (VMs)} run a full guest operating system on virtual hardware provided by a hypervisor. On Linux, KVM turns the kernel into a hypervisor, and QEMU~\cite{bellardQEMUFastPortable2005} emulates the devices the guest needs. This provides strong isolation, at the cost of more software layers and of transitions between host and guest.

\textbf{Unikernels}~\cite{madhavapeddyUnikernelsLibraryOperating2013, kuenzerUnikraftFastSpecialized2021, kivityOSvOptimizingOperating} build an application together with only the kernel parts it uses, producing a small single-address-space image that runs directly on a hypervisor. They therefore use less memory, boot faster, and expose a smaller attack surface, a good fit for short, single-purpose serverless functions that still need isolation and fast startup~\cite{caddenSEUSSSkipRedundant2020a, moebiusAreUnikernelsReady2024}. Since they require hardware virtualization, their efficiency also depends on the hypervisor they run on.

\textbf{MicroVMs} are a lightweight VM model designed for serverless environments. Firecracker~\cite{firecracker}, built by AWS for Lambda, emulates only the devices a function needs: it keeps network and block devices, but drops BIOS, PCI, USB, and GPU support to reduce memory footprint and startup time. MicroVMs thus provide the isolation of a VM while approaching the lightness of a container~\cite{firecracker, goethalsFunctionalPerformanceBenchmark2022}.

\subsection{Serverless Execution Environments}
Researchers have studied the cost of the execution environment primarily using a single metric: cold start time. 
Several systems try to remove it.
SEUSS~\cite{caddenSEUSSSkipRedundant2020a} runs functions inside unikernels and caches a snapshot of the whole memory and register state of a running function, so that a new instance skips three steps at once: booting the unikernel, starting the language runtime, and importing and compiling the function code. 
Catalyzer~\cite{duCatalyzerSubmillisecondStartup2020} applies the same idea to gVisor sandboxes and achieves sub-millisecond startup by restoring a checkpoint rather than booting.
Ustiugov~\emph{et~al.}~\cite{ustiugovBenchmarkingAnalysisOptimization2021} study why snapshot restoration is still slow, and show that prefetching the pages touched during the first invocation reduces cold start latency by 3.7$\times$. 
These works show that the environment incurs a real cost, but they optimize it rather than compare technologies, and none report energy.
 
More closely related to our work, a few studies directly compare virtualization technologies. 
Goethals~\emph{et~al.}~\cite{goethalsFunctionalPerformanceBenchmark2022} benchmark containers, Firecracker microVMs, and OSv~\cite{osv} unikernels for edge microservices, on x86 and on a Raspberry Pi, looking at boot time, memory use, network performance, and toolchain maturity. 
They find that Docker containers deliver reliable performance with low memory usage, while microVMs provide better isolation, and OSv unikernels boot extremely fast and can outperform Docker containers. 
Moebius~\emph{et~al.}~\cite{moebiusAreUnikernelsReady2024} study unikernels (Nanos, OSv) as FaaS sandboxes against Docker, gVisor, and a Linux microVM. 
They show that unikernels can reduce cold-start CPU cost by up to 8.5$\times$ compared to a microVM and 2.5$\times$ compared to gVisor, but also that with a heavy runtime, such as Node.js, starting the application costs much more than starting the sandbox. 
Dinh-Tuan~\cite{dinhtuanUnikernelsVsContainers2025} compares Docker and Nanos under strict memory limits, with a Go application (ahead-of-time compiled) and a Node.js application (JIT). 
Unikernels boot faster and outperform  Docker for Go, but for Node.js there is a crossover: below a certain memory limit, Docker stays stable while Nanos degrades sharply. 
Unlike these three studies, which report latency, memory, or CPU cost but not energy, our work measures energy directly through hardware RAPL counters and decomposes it phase by phase (boot, compute, teardown), across a broader set of experiments. 

\subsection{Energy efficiency in serverless computing}
\textbf{Measuring and attributing energy.}  
Rehman~\emph{et~al.}~\cite{faasmeter} present FaasMeter, a FaaS control plane where energy monitoring, accounting, and pricing are first-class operations. 
Because system-level power readings are noisy and conflate all functions, they disaggregate them using power models, Kalman filters, and Shapley values to obtain per-function footprints accurate to about 1\%, including the control plane itself. 
Basu Roy~\emph{et~al.}~\cite{royHiddenCarbonFootprint2024} apply a similar accounting effort to carbon, and show that a large part of the footprint of a serverless platform never appears in the execution time billed to the user. 
In the same spirit, Da Silva~\emph{et~al.} investigate the performance of HPC workflows and show that serverless computing may be an option for reducing its energy consumption~\cite{dasilva-sc24}, and also adapt the \emph{Power Usage Effectiveness} (PUE) metric to serverless platforms to separate the power that reaches the functions from the power spent by the platform around them~\cite{dasilvaAssessingPowerUsage2025}.

\textbf{Characterizing and reducing energy.}  
Stojkovic~\emph{et~al.}~\cite{stojkovicEcoFaaSRethinkingDesign2024} characterize serverless environments and observe that functions run in opaque sandboxes, are idle for a large share of their invocation time, switch context often, and are co-located dynamically---a combination that defeats classic energy-management schemes; they then redesign the runtime accordingly. 
This idleness is consistent with the production traces analyzed by Shahrad~\emph{et~al.}~\cite{shahradServerlessWildCharacterizing2020}, in which most functions are short and invoked infrequently. 
Other work reduces the waste rather than measuring it: MXFaaS~\cite{stojkovicMXFaaSResourceSharing2023} multiplexes processor cycles, I/O bandwidth, and memory state between concurrent invocations of the same function; EneX~\cite{rastegarEneXEnergyAwareExecution2024} schedules invocations under an energy budget; and Da Silva~\emph{et~al.}~\cite{dasilva-acmsac25} schedule serverless functions using a multi-objective allocation policy that optimizes execution time and energy efficiency at the same time, however within an offline approach, usable only for batches of functions.

\subsection{Programming Languages as Energy Factor}
Pereira et al.~\cite{pereiraEnergyEfficiencyProgramming2017} measured the energy consumption, execution time, and memory usage of 27 programming languages on workloads from the Computer Language Benchmarks Game (CLBG)~\cite{clbg_debian}. 
They show that the programming language is a major energy factor: compiled languages such as C and Rust consume far less energy than interpreted languages such as Python, and the fastest language is not always the most energy-efficient. 
Their study is the closest to ours along the language axis, but it runs on bare metal only, with each language's default runtime, and outside any serverless lifecycle: no boot, no teardown, and no network I/O.

\section{A framework for benchmarking the energy efficiency of functions}\label{sec:framework}
In this section, we describe our methodology to evaluate the energy consumption of serverless functions and their execution environments. 
By \emph{execution environment}, we mean the set of technologies used to execute a function implemented in source code.
We call \emph{worker} the executable that wraps a function for a given language-runtime configuration: it receives a job, runs the algorithm, stores the result, and exits. 
Each worker runs inside one \emph{sandbox}, the isolation unit provided by the execution environment. For instance, a sandbox can be a Linux process on a bare-metal host.
We selected several virtualization technologies and workloads, written in different programming languages, to study their impact on energy consumption. 
In the remainder of this section, we recall and detail our research questions (in Section~\ref{sec:research-questions}), our experimental methodology (in Section~\ref{sec:methodology}), and our design of experiments (in Section~\ref{sec:doe}).

\subsection{Research questions}\label{sec:research-questions}
We refine here the three research questions of Section~\ref{sec:intro}: 

\textbf{RQ1} asks whether the most popular programming languages for serverless functions are the most energy-efficient.
Pereira~\emph{et al.} ranked languages by energy on bare metal~\cite{pereiraEnergyEfficiencyProgramming2017}, but in a serverless deployment a function's energy also includes lifecycle overheads, such as environment initialization and teardown, and may depend on the execution environment.
We therefore compare language--runtime configurations across execution environments, and ask whether the serverless lifecycle changes the ranking observed on bare metal.

\textbf{RQ2} asks which factors most affect the energy efficiency of serverless functions.
We study the container engine, base image, hypervisor, unikernel, and function duration, and analyze each lifecycle phase separately, to identify not only which environments consume more energy but also which phases cause the differences.

\textbf{RQ3} asks which configurations are the most energy-efficient to deploy.
Since boot and teardown are paid once per invocation while computation energy grows with the work, we express the lifecycle cost as the computation time needed to amortize it, and compare configurations across languages and environments.

\subsection{An experimental methodology}\label{sec:methodology}

\subsubsection{\textbf{Virtualization technologies}}\hfill
\label{selected-virtualization-technologies}

To evaluate the impact of virtualization technologies, we select several representative approaches that are either used in production and open-source FaaS platforms, such as containers in Knative, OpenFaaS, and Apache OpenWhisk~\cite{knative,openfaas,openwhisk} and Firecracker in AWS Lambda~\cite{firecracker}, or proposed as FaaS sandboxes in the literature, such as unikernels~\cite{caddenSEUSSSkipRedundant2020a,moebiusAreUnikernelsReady2024}. 

We use \textbf{bare-metal} execution as the reference configuration. 
Applications are executed directly on the host operating system, without hardware virtualization or containerization. 
This configuration provides a baseline and lets us quantify the overhead introduced by different abstraction layers.

For \textbf{containerization}, we selected Docker~\cite{docker} and Podman~\cite{podman}. 
Both rely on Linux namespaces and cgroups, use the runc runtime, and share the host kernel. 
Docker is widely used in industry and in cloud platforms, while Podman is a daemonless alternative, often chosen for its security features.
For each \textbf{container engine}, we built two images that differ only in their base layer: a standard Debian image linked against glibc, and an Alpine image linked against musl. 
Small images are popular in serverless deployments because downloading the image is a large part of container startup time~\cite{harterSlackerFastDistribution}. 
In our setup, however, we store all images on the server before the experiments, so we download no images during a run. 
Therefore, we measure only the effect of the C library (glibc vs.\ musl), not image size.

For the \textbf{unikernel} approach, we chose Nanos and Unikraft, two open-source projects that adopt different strategies~\cite{kuenzerUnikraftFastSpecialized2021}. 
Nanos aims to run existing applications in a unikernel environment with minimal adaptation, whereas Unikraft uses a modular architecture that lets you select only the system components the application needs, promoting high specialization. 
This choice enables us to cover two design strategies: application compatibility and minimalism.
Because unikernels require hardware virtualization, we evaluated two \textbf{hypervisors}: QEMU with KVM, representing traditional hardware virtualization, and Firecracker, which implements a minimalist microVM model designed for serverless environments. 
This comparison allows us to assess whether the choice of the hypervisor, general-purpose or specialized, affects the energy efficiency and performance of unikernels.


\subsubsection{\textbf{Workloads, adaptations and input sizes}}\hfill
\label{sec:workloads}
Existing serverless benchmark suites, such as SeBS~\cite{copik2021sebs} and FunctionBench~\cite{kimFunctionBenchSuiteWorkloads2019}, provide realistic application workloads. 
However, their functions are typically implemented in only one or a few programming languages, predominantly Python and JavaScript, and may rely on platform-specific APIs. 
Consequently, they do not generally allow the same algorithm to be evaluated across multiple programming languages and language-runtime configurations, which is essential for addressing RQ1.
We therefore select our workloads from the \textbf{\emph{Computer Language Benchmarks Game} (CLBG)}~\cite{clbg_debian}, which Pereira~\emph{et~al.}~\cite{pereiraEnergyEfficiencyProgramming2017} also use, allowing us to compare our results with theirs.
The CLBG provides implementations of the same algorithms in multiple languages, keeping the computational task fixed while varying the language-runtime configuration. We select ten programs, presented in Table~\ref{tab:workloads}, that vary among CPU-, Memory-, I/O-intensive, and library-bound workloads. 
Unlike realistic applications, which usually exercise several resources at once, ours mainly stresses one resource at a time, letting us identify which computation type each layer affects most.
For instance, a runtime may be efficient on CPU-bound workloads but less so on memory-intensive ones. 
Finally, we add one workload of our own, called \emph{noop}, which isolates the execution environment's energy overhead. 

The original programs read their parameters from the command line and write to standard output, whereas serverless functions are triggered by events and exchange data with external services; we therefore \textbf{adapt their inputs and outputs}, without changing the algorithms.
Each job is announced by a message in a RabbitMQ~\cite{rabbitmq} queue, which points to a job request stored in MinIO~\cite{minio}, an S3-compatible object store~\cite{amazons3}.
The worker claims the message, reads the request, and, for \emph{k-nucleotide}, \emph{regex-redux}, and \emph{reverse-complement}, downloads an input dataset generated once with \emph{fasta}.
Instead of writing to standard output, it uploads its result to the object store under \texttt{results/}, together with a metadata object recording the execution status and the start and end timestamps of the computation, which lets us verify that each run completed as expected.

Serverless functions can run from milliseconds to several minutes; however, the relative impact of execution-environment overhead varies across this range.
To study this effect, we execute each workload with \textbf{three input parameters}. 
We select the corresponding parameters using Rust on bare-metal (our fastest configuration) as the reference configuration, such that the core computation lasts approximately \textbf{100\,ms}, \textbf{1\,s}, and \textbf{10\,s}. 
We then use the same input parameters across all other languages and environments, ensuring that every configuration executes the same algorithmic workload. 
These three durations follow the production traces of Shahrad \emph{et~al.}~\cite{shahradServerlessWildCharacterizing2020}: about 10\% of functions run under 100\,ms, half under 1\,s, and 90\% under 10\,s.

\begin{table}[htbp]
\caption{Our set of workloads and their characteristics. $\blacksquare$: the workload reads its input or writes its output through the object store; $\square$: it does not.}
\label{tab:workloads}
\centering
\scriptsize
\begin{tabular}{@{}lllc}
\toprule
\bf Workload & \bf Instructions & \bf Workload & \bf \makecell{Data\\intensive} \\
\midrule
\em mandelbrot         & floating point, small loops   & CPU-intensive & $\blacksquare$ \\
\em n-body             & floating point, small data    & CPU-intensive & $\square$ \\
\em spectral-norm      & floating point, nested loops  & CPU-intensive & $\square$ \\
\em fannkuch-redux     & integers, arrays, branches    & CPU-intensive & $\square$ \\
\midrule
\em binary-trees       & allocator, garbage collector  & Memory-intensive & $\square$ \\
\em k-nucleotide       & hash tables, strings          & Memory-intensive & $\blacksquare$ \\
\midrule
\em pidigits           & big-number library            & Library-bound & $\square$ \\
\em regex-redux        & regular expression engine     & Library-bound & $\blacksquare$ \\
\midrule
\em fasta              & data generation, output       & I/O-intensive & $\blacksquare$ \\
\em rev-comp           & string handling, I/O          & I/O-intensive & $\blacksquare$ \\
\midrule
{\em noop} (control)   & nothing ($\mathrm{sleep}(T)$, T = $N$ ms) & None & $\square$ \\
\bottomrule
\end{tabular}
\end{table}


\subsubsection{\textbf{Programming languages}}
\label{sec:languages} 
To study the impact of programming languages across execution environments, 
we select several of the most widely used on serverless platforms~\cite{datadog2024serverless, eismann2021serverless,eskandaniWonderlessDatasetServerless2021}. 
For every language, we use the fastest runtime available for our workloads.
This choice avoids blaming a language for a slow runtime and reflects what a performance-oriented serverless provider could actually deploy.

\textbf{Rust} (rustc 1.100.0-nightly) and \textbf{Go} (v1.25.10): they are compiled ahead of time to native binaries.
They do not need a separate runtime such as a JVM or an interpreter. 
Go's garbage collector and scheduler are linked into the binary.
Both produce a single binary with a small memory footprint, which is simple to package in a container or a unikernel image. 
We use Rust as our language baseline because it often ranks among the most energy-efficient languages on CLBG workloads~\cite{pereiraEnergyEfficiencyProgramming2017}.
We use two baselines: bare metal for execution environments and Rust for language-runtime configurations.
\textbf{Java} (GraalVM Native Image v25.02, Maven v3.9.11): we do not use a classic JVM. 
We compile the Java workers ahead of time with GraalVM Native Image. 
This removes JVM startup and JIT warm-up, known overheads for Java in serverless environments~\cite{carreiraWarmHotStarts2021}. 
However, the native binary may not reach the same peak performance as a warmed-up JVM. 
Frameworks such as Quarkus~\cite{quarkus} and Micronaut~\cite{micronaut} support native compilation for serverless deployments.
\textbf{JavaScript (Bun v1.3.14, fallback: Node.js v24.18)}: Bun runs JavaScript on JavaScriptCore and starts much faster than Node.js (V8)~\cite{merelo-guervosAnalysisEnergyConsumption2023}. 
When Bun cannot be deployed, the worker falls back to Node.js.
\textbf{Python (PyPy v3.11, fallback: CPython v3.12)}: PyPy is a Python implementation with a just-in-time (JIT) compiler. 
It is faster than CPython, the reference interpreter, on most benchmarks, including CLBG programs~\cite{pypy, clbg_debian}. 
When PyPy cannot be deployed, the worker falls back to CPython.

These choices cover three execution models: ahead-of-time compilation (Rust, Go, Java Native Image), JIT compilation (Bun on JavaScriptCore, Node.js on V8, PyPy), and bytecode interpretation (CPython). 
The execution model influences startup cost, memory behavior, and how close the code runs to native speed. 
These properties are the most likely to interact with the execution environment.
Runtime availability limits the configurations we can evaluate because some runtimes are not available in every execution environment. 
PyPy does not provide an official musl build, so Alpine images lack Python. 
Neither Bun nor PyPy runs on Nanos and Unikraft, where workers fall back to Node.js and CPython. 
Finally, GraalVM Native Image cannot statically link the gmp and pcre2 libraries against musl, so the Java versions of pidigits and regex-redux are absent from the two Alpine images. 
Table~\ref{tab:runtime-matrix} gives the implementation used in each environment.
Strictly speaking, we therefore study language-runtime configurations rather than languages alone; in the rest of the paper, we use the language name (e.g.,Python) to refer to its configuration when the runtime is clear from context.

\begin{table}[htbp]
\caption{Implementation used for each language and execution environment.}
\label{tab:runtime-matrix}
\centering
\scriptsize
\begin{tabular}{@{}lccccc@{}}
\toprule
\textbf{Environment} &
\textbf{Rust} &
\textbf{Go} &
\textbf{Java} &
\textbf{Python} &
\textbf{JavaScript} \\
\midrule
Bare metal             & native & native & GraalVM & PyPy               & Bun \\
Docker (std.)          & native & native & GraalVM & PyPy               & Bun \\
Docker (Alpine)        & native & native & GraalVM & -                  & Bun \\
Podman (std.)          & native & native & GraalVM & PyPy               & Bun \\
Podman (Alpine)        & native & native & GraalVM & -                  & Bun \\
Nanos (QEMU/KVM)       & native & native & GraalVM & CPython$^\dagger$  & Node.js$^\dagger$ \\
Nanos (Firecracker)    & native & native & GraalVM & CPython$^\dagger$  & Node.js$^\dagger$ \\
Unikraft (QEMU/KVM)    & native & native & GraalVM & CPython$^\dagger$  & Node.js$^\dagger$ \\
Unikraft (Firecracker) & native & native & GraalVM & CPython$^\dagger$  & Node.js$^\dagger$ \\
\bottomrule
\end{tabular}

\footnotesize
$^\dagger$ Neither Bun nor PyPy could be deployed on the unikernel technologies (Nanos, Unikraft), Node.js and CPython were used instead.
\end{table}

\subsubsection{Workers' life cycle.} The worker emits a start marker, connects to the object store and the queue, and polls the queue: it claims a job, runs the algorithm, uploads the result, and exits once it has been idle longer than a threshold. We set this threshold to 0, so each worker processes exactly one job, and every measurement is a cold start; we leave warm batches to future work. Markers split each run into five parts: sandbox creation, connection setup and job retrieval, computation, result upload, and sandbox destruction. Using an actual message broker and object store ensures that the worker performs the network operations of a real serverless function.

\subsubsection{\textbf{Testbed Setup}}
\textbf{Energy measurement.} We measure energy with Joule Profiler~\cite{Woirhaye_Joule_Profiler_A_2026}, a lightweight tool that reads hardware energy counters and, unlike system-level tools such as Alumet~\cite{alumet}, PowerAPI~\cite{powerapi}, or Scaphandre~\cite{scaphandre}, attributes energy to individual execution phases rather than reporting a single aggregate value. 
Phase boundaries are defined by markers emitted by the worker, requiring only minimal code changes. Joule Profiler reads Intel's Running Average Power Limit (RAPL) interface through \texttt{perf\_event}, for the package and DRAM domains.

\textbf{Execution phases.} 
Joule\,Profiler divides each run into three main phases: 
1) the \emph{boot} phase, which runs from the start of the environment until the worker is ready to take a function; 
2) the \emph{core computation} phase, which corresponds to the execution of our workloads; 
3) the \emph{teardown} phase, which corresponds to the complete destruction of the environment. 
We use these phases to distinguish the energy associated with the execution environment from that associated with the application computation, as their relative contributions may vary with workload size and execution time.
 
\textbf{CPU pinning.} 
Our server has two CPU sockets, and RAPL reports one energy counter per socket. 
If a run could migrate from one socket to the other, its energy would be split across two counters, and each counter would also include unrelated activity. 
To attribute the energy of a run to a single counter, we pin each run to one socket and report only the RAPL package energy of that socket.
Within the selected socket, cold-start experiments use a single CPU core so that all languages get the same degree of parallelism. Each workload runs in a dedicated \texttt{systemd} slice pinned to that socket and its NUMA memory node (\texttt{AllowedCPUs}, \texttt{AllowedMemoryNodes}), with a memory cap and swap disabled (\texttt{MemoryMax}, \texttt{MemorySwapMax=0}); the \texttt{cgroup v2} hierarchy applies these limits uniformly to containers, virtual machines, and processes.
 
\textbf{Hardware configuration.} 
We conduct all experiments on a dedicated bare-metal server.
A dedicated testbed is important because energy measurements can be affected by the interference from the workloads of other users. The node is equipped with two Intel Xeon Gold 6126 processors (12 cores each, x86\_64) and 192\,GB of RAM, and runs Debian 12.11 (kernel 6.1.0-44), deployed using kadeploy~\cite{jeanvoineKadeploy3EfficientScalable2013}. 



\subsection{Design of experiments}
\label{sec:doe}
We measure the impact of the execution environment, the language runtime, and the workload on energy. 
Table~\ref{tab:doe} summarizes our design of experiments. 
Each configuration is defined as a combination of execution environment, language-runtime, workload, and input size: $[\textit{environment} \times \textit{language} \times \textit{workload} \times \textit{input size}]$. 
The full cross-product would yield 693 configurations per input size ($9 \times 7 \times 11$), but 114 of them cannot be deployed, because some runtimes or native libraries are not available in every execution environment (Section~\ref{sec:languages}, Table~\ref{tab:runtime-matrix}). 
This leaves 579 valid configurations at the shortest input size. Each configuration constitutes a separate run, which we repeat $N$ times. 
The number of repetitions depends on the input size: $N=60$ for the 100\,ms run, $N=40$ for the 1\,s run, and $N=20$ for the 10\,s run. 
We use more repetitions for shorter runs, because fixed-size perturbations (timer interrupts, background activity, marker latency) weigh more in a short measurement window, which makes short runs noisier in relative terms.

RAPL measures the energy consumption of the entire socket rather than that of an individual process. 
We therefore define the energy consumed during a phase as the energy measured by the RAPL package counter of the CPU socket on which the execution is pinned during that phase.
This measurement includes the energy consumed by the worker process, the container engine, or the virtual machine monitor and guest kernel. 
This is the relevant scope for a provider, which pays the energy of the entire sandbox rather than paying solely the energy of the code running inside it.

\begin{table}[ht]
  \caption{Design of experiments.}
  \label{tab:doe}
  \centering
  \footnotesize
  \begin{tabular}{lrrr}
    \toprule
    \textbf{Parameters} & \multicolumn{3}{c}{\textbf{Values}} \\
    \midrule
    \textbf{Input size} (see Section~\ref{sec:workloads}) & 100\,ms & 1\,s & 10\,s \\
    \textbf{Environments} (see Table~\ref{tab:runtime-matrix})  & 9 & 9 & 9 \\
    \textbf{Workloads} (see Table~\ref{tab:workloads})          & 11 & 11 & 11 \\
    \textbf{Language-runtime config.} (see Section~\ref{sec:workloads})                & 7 & 6 & 4 \\
    \midrule
    \textbf{Valid configurations$^\dagger$}                           & 579 & 480 & 342 \\
    \textbf{Total of experiments} & \multicolumn{3}{c}{1401} \\
    \bottomrule
  \end{tabular}
  
\footnotesize
 The slowest language-runtime configurations are dropped for higher input sizes because a single run becomes too long to repeat at scale for little added information (see Section~\ref{sec:doe}).
\end{table}
\vspace{-0.5cm}
%

\section{Experimental results}\label{sec:results}
%
This section answers the three research questions of \S\ref{sec:research-questions}.
\textbf{RQ1} (\S\ref{sec:rq1}) asks whether the languages serverless developers actually use are also the most energy-efficient, and whether the serverless lifecycle reshuffles the ranking that prior work established on bare metal~\cite{pereiraEnergyEfficiencyProgramming2017}.
\textbf{RQ2} (\S\ref{sec:rq2}) examines what else drives energy consumption: the execution environment, the layers within it (container engine, base image, hypervisor, unikernel), and the duration of the function.
\textbf{RQ3} (\S\ref{sec:rq3}) combines these answers into a decision procedure for developers and platform designers.

Unless stated otherwise, we report the cold-start energy $E_{cold}$: the RAPL package energy consumed during the \emph{boot}, \emph{core computation}, and \emph{teardown} phases.
\emph{Harness} phases (queue and object-store connection, job retrieval, result upload) are measured separately and not included in $E_{cold}$.
DRAM adds about $16\%$ to the package energy but leaves every ranking unchanged, so we omit it.

Our measurements are stable. 
Across the 579 configurations at the 100\,ms input size, the shortest and therefore the noisiest, the median coefficient of variation over the 60 repetitions is 6.4\%, which puts the 95\% confidence interval of each configuration at about $\pm$1.6\%. 
No configuration exhibits a bimodal distribution.



\subsection{RQ1: Are the most popular programming languages for implementing serverless functions the most energy efficient?}\label{sec:rq1}

\subsubsection{The popular languages are the expensive ones}\label{sec:rq1-popular-languages}
\begin{figure}[htbp]
    \centering
    \includegraphics[width=\columnwidth]{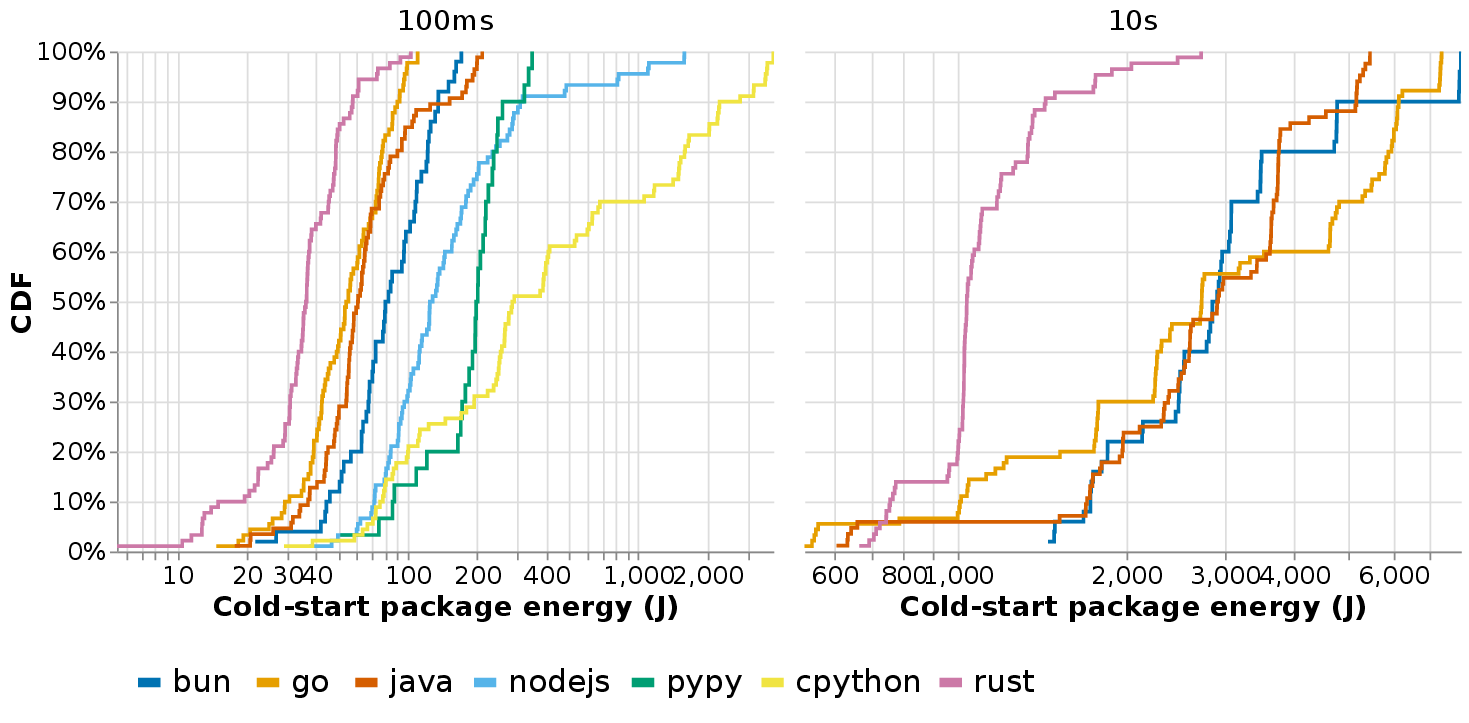}
    \caption{Empirical CDF of cold-start energy per language-runtime configuration, at 100\,ms and 10\,s (log $x$-axis). Curves do not pool the same environments (Table~\ref{tab:runtime-matrix}). The ranking stays stable across durations despite higher absolute energy at 10\,s (Table~\ref{tab:lang}).}
\label{fig:lang-cdf}
\end{figure}

Python and JavaScript dominate production serverless deployments~\cite{datadog2024serverless, eismann2021serverless, eskandaniWonderlessDatasetServerless2021} and their default runtimes sit at the expensive end of our measurements. At 100\,ms, CPython is the rightmost curve of Figure~\ref{fig:lang-cdf} and Node.js lies to the right of Rust, Go, Java and Bun: CPython consumes $11.4\times$ the cold-start energy of Rust, and Node.js $4.3\times$ (see Table~\ref{tab:lang}).

The Friedman test confirms this language effect ($\chi^2(4) = 282.0$, $p < 10^{-50}$, $n = 86$ complete blocks, where a block is one workload-environment pair). 
The test covers the five configurations deployable in every environment. The ranking is nearly the same in every block (Kendall's $W = 0.82$; Figure~\ref{fig:rank-stability}), but it is a partial order: the five configurations form three groups rather than five distinct ranks. 
\emph{Group~1} contains Rust alone (mean rank 1.15). 
\emph{Group~2} includes the other ahead-of-time compiled configurations, Go (2.20) and Java with GraalVM Native Image (2.83), whose order changes across workloads. 
\emph{Group~3} contains Node.js (4.17) and CPython (4.65). 
Bun and PyPy cannot be deployed everywhere and are therefore outside this test; in the pooled comparison of Table~\ref{tab:lang}, Bun falls between Groups~2 and~3, and PyPy between Node.js and CPython.

The ranking also holds for longer functions: at 10\,s, the four configurations we measured (Rust, Go, Java, Bun) share the same ranks as at 100\,ms (see Table~\ref{tab:lang}), although each configuration consumes more energy in absolute terms.

\begin{table}[htbp]
    \scriptsize
    \caption{Cold-start energy of each language-runtime configuration relative to a reference, per input size (geometric mean, all ratios significant, $p_H < 0.01$). Groups come from the Friedman test (\S\ref{sec:rq1}); a dash means the configuration was not measured (Table~\ref{tab:runtime-matrix}).}
    \label{tab:lang}
    \centering
    \resizebox{\linewidth}{!}{ %
    \begin{tabular}{lcc rrr}
        \toprule
        \bf Configuration & \bf Reference & \bf Group & \bf 100\,ms & \bf 1\,s & \bf 10\,s \\
        \midrule
        Go                   & Rust    & 2  & $\times$1.55  & $\times$2.23 & $\times$2.43 \\
        Java (native image)  & Rust    & 2  & $\times$1.87  & $\times$2.32 & $\times$2.51 \\
        Bun (JavaScript)     & Rust    & -- & $\times$2.33  & $\times$2.66 & $\times$2.85 \\
        Node.js (JavaScript) & Rust    & 3  & $\times$4.32  & $\times$4.18 & --           \\
        PyPy (Python)        & Rust    & -- & $\times$6.08  & $\times$8.43 & --           \\
        CPython (Python)     & Rust    & 3  & $\times$11.42 & --           & --           \\
        \midrule
        Bun                  & Node.js & -- & $\times$0.69  & $\times$0.82 & --           \\
        PyPy                 & CPython & -- & $\times$0.52  & --           & --           \\
        \bottomrule
    \end{tabular}
  }
\end{table}

\subsubsection{The runtime is a first-class parameter, not an implementation detail}
The programming language is not the whole story. 
For the \emph{same source code}, switching runtimes changes energy by a fifth to a half. All ratios reported below are taken from Table~\ref{tab:lang}, whose last two rows give the same-language comparisons. At 100\,ms, Bun uses 31\% less cold-start energy than Node.js (50 matched pairs, $p_H = 2.8 \times 10^{-14}$), and PyPy 48\% less than CPython (30 pairs, $p_H = 1.5 \times 10^{-3}$). 
The advantage of Bun stems largely from its faster start-up: it decreases to 18\% at 1\,s input size.
Compiled ahead of time with GraalVM Native Image, Java consumes $1.87\times$ the energy of Rust, within $21\%$ of Go and less than Bun, and therefore belongs to the group of compiled languages, where a normal JVM start-up never places it.
These gains are as large as a change of the whole execution environment, and larger than a change of any single layer of it: switching from Docker to Podman saves 14\%, and from QEMU to Firecracker 5 to 18\% (see Table~\ref{tab:contrasts}), while the largest gap between two virtualized environments is 56\%. 
A Python or JavaScript function therefore does not need to be rewritten to consume less energy: it only needs a faster runtime, although such a runtime is not always available (see Table~\ref{tab:runtime-matrix}).

\subsubsection{Does the serverless lifecycle change the bare-metal ranking?}

\begin{figure}[htbp]
    \centering
    \includegraphics[width=\columnwidth]{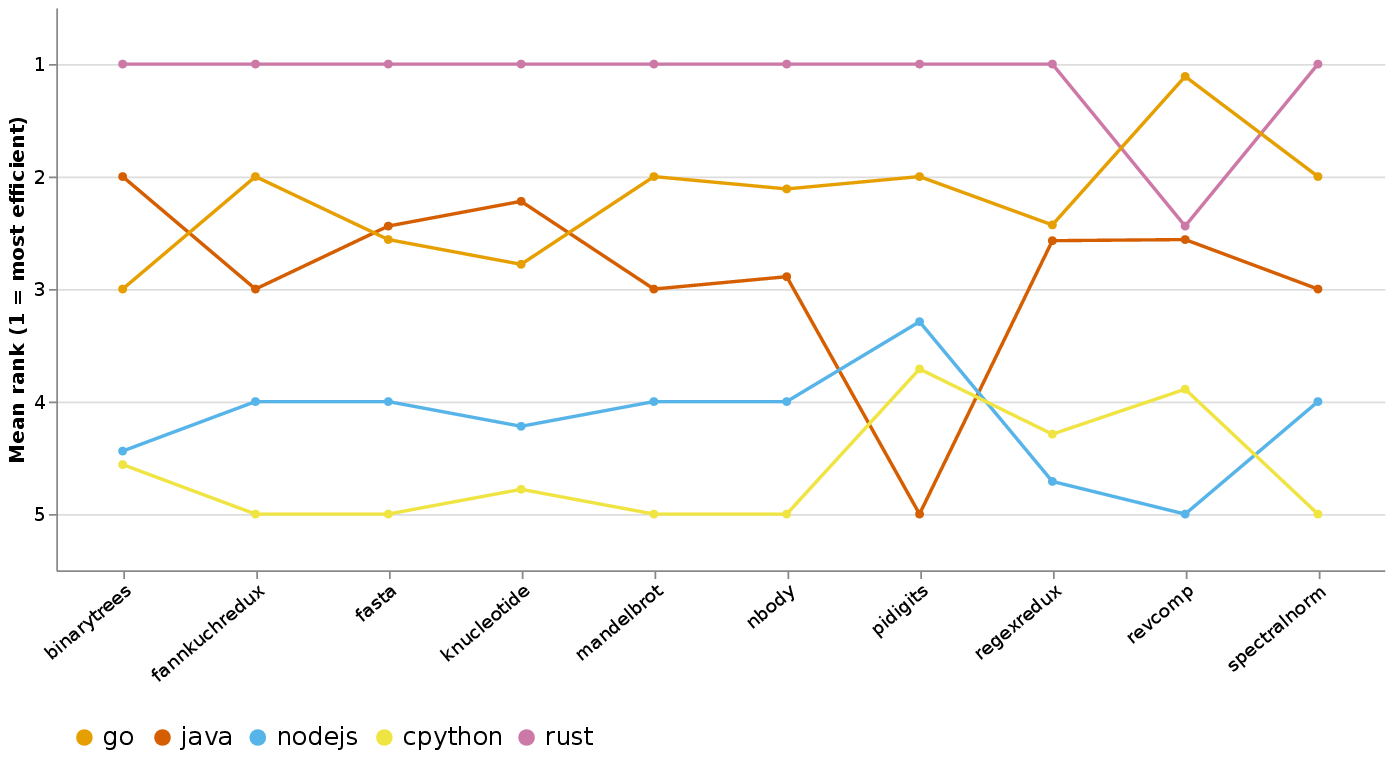}
    \caption{Average ranking of each language-runtime configuration per workload (1 = lowest cold-start energy, 100\,ms). Bun and PyPy excluded (not deployable everywhere, Table~\ref{tab:runtime-matrix}). Rankings are stable (Kendall's $W \geq 0.89$); the few swaps involve native libraries (\emph{pidigits}, \emph{regex-redux}, \emph{reverse-complement}).}
    \label{fig:rank-stability}
\end{figure}

Prior work established programming-language energy rankings on bare-metal, based on computation alone~\cite{pereiraEnergyEfficiencyProgramming2017}. 
A serverless invocation adds a boot, a teardown, and network I/O around that computation, all of which are largely language-independent.

\textbf{The ranking survives.} 
We repeat the language test workload by workload, blocking by environment, across all 9 environments, and Kendall's $W$ stays between 0.89 and 1.00 everywhere: the ranking reported in Figure~\ref{fig:rank-stability} barely changes compared to the results reported~\cite{pereiraEnergyEfficiencyProgramming2017}. 
The same rank holds inside each of the nine environments, in containers and unikernels alike. 
The test covers the five configurations deployable everywhere; Bun and PyPy, which cannot be deployed on the unikernels (see Table~\ref{tab:runtime-matrix}), keep their rank in the pooled comparison of Table~\ref{tab:lang}.

The few visible swaps all involve a native library, and they coincide with the lowest values of $W$.  
On \emph{pidigits}, a workload dominated by a big-integer library, CPython ranks fourth, since its integer arithmetic executes as optimized C code, whereas Java ranks last, since its \texttt{BigInteger} is implemented in plain Java.
On \emph{regex-redux}, CPython surpasses Node.js for the same reason. 
On \emph{reverse-complement}, Go ranks first, ahead of Rust, and $W$ reaches its minimum of 0.89. 
These swaps reflect the maturity of a language's ecosystem rather than the performance of the language itself, and they illustrate the limits of any single global ranking.

\textbf{But the gap between languages widens with input size.}
The serverless lifecycle does not change the ranking, but it does change how far apart the languages are.
Boot and teardown are a fixed cost per invocation, paid identically by a fast and a slow language, so they dilute the fast one's advantage: the shorter the computation, the more this fixed cost dominates $E_{cold}$, and the closer every language's total energy is pulled toward it.
As input size grows, the fixed cost becomes negligible relative to the computation, and the gap widens to reflect the languages' actual difference in efficiency.
Table~\ref{tab:lang} shows this effect for most configurations: relative to Rust, the ratio grows from $\times1.55$ at 100\,ms to $\times2.43$ at 10\,s for Go, from $\times1.87$ to $\times2.51$ for Java, from $\times2.33$ to $\times2.85$ for Bun, and from $\times6.08$ at 100\,ms to $\times8.43$ at 1\,s for PyPy.
Node.js is the only exception ($\times4.32$ at 100\,ms, $\times4.18$ at 1\,s): its own boot is expensive, so a longer computation dilutes \emph{its} fixed cost too, which compensates for its slower computation and narrows the gap instead of widening it.
In other words, the ranking observed on bare metal still holds in a serverless deployment, but for most runtimes the gap it reveals only fully emerges once the function runs long enough for the fixed cost to stop masking it.
\S\ref{sec:rq3} quantifies how long a function must run to reach that point.


\begin{tcolorbox}[boxrule=0.4pt, breakable]
\textbf{Answer to RQ1.}
No: the most popular languages are the least energy-efficient, by up to $\times11.4$ against Rust, and this ranking holds across environments and almost every workload.
The lifecycle does not reorder languages; it only narrows the gaps for short functions.
The only lever that helps a popular language without rewriting it is the runtime: Bun and PyPy save 31\% and 48\%, respectively.
\end{tcolorbox}


\subsection{RQ2: What are the key factors impacting the energy efficiency of serverless functions?}\label{sec:rq2}
Once the language is selected, three factors remain: the environment, its layers, and the function's duration.

\subsubsection{The execution environment}\label{sec:rq2-env}
\begin{figure}[htbp]
  \centering
  \includegraphics[width=\linewidth]{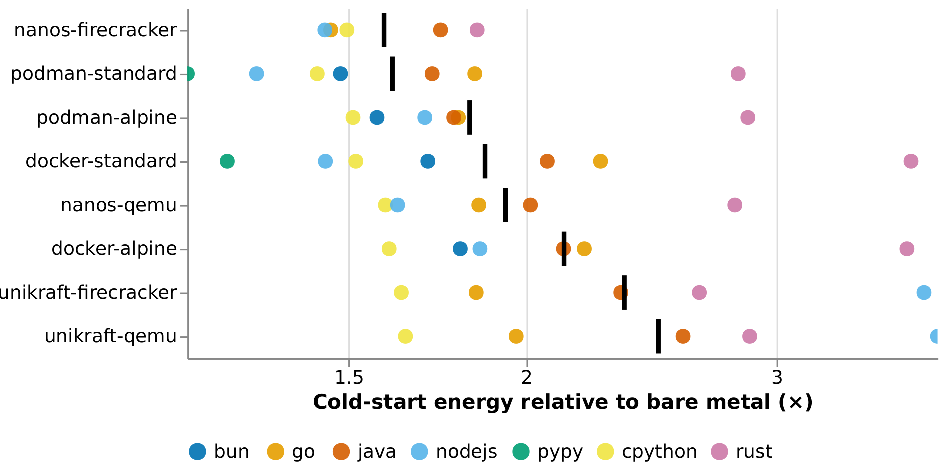}
  \caption{Cold-start energy of each virtualized environment relative to bare metal, at 100\,ms (geometric mean over matched workloads). Each point is a language-runtime configuration; the black tick is the mean. Environments are sorted by this mean; containers and unikernels are mixed, and the rank changes with the runtime.}
  \label{fig:env-vs-bm}
\end{figure}

The Friedman test also detects an environment effect ($\chi^2(8) = 213.9$, $p = 7.60 \times 10^{-42}$, $n = 48$ blocks, one per workload and language-runtime pair). 
But this effect differs in nature from the language effect. The environment ranking is much less consistent across blocks ($W = 0.557$, against $0.820$ for the language, \S\ref{sec:rq1}), and its magnitude is much smaller.
Figure~\ref{fig:env-vs-bm} shows the cost of each environment relative to bare-metal at the 100\,ms input size. 
Bare-metal ranks first in every block (mean rank 1.00), and every virtualized environment adds between 59\% (Nanos on Firecracker) and 148\% (Unikraft on QEMU) to $E_{cold}$. 
Behind bare-metal, the eight virtualized environments fall into two families: containers (Docker and Podman, each with a standard and an Alpine image) and unikernels (Nanos and Unikraft, each on Firecracker and on QEMU). The two families are interleaved in the ranking: Nanos on Firecracker ranks first (2.94), followed by Podman with a standard image (4.08), and Unikraft on QEMU ranks last (7.17).

%
The two families do not separate. 
The best unikernel, Nanos on Firecracker, consumes 9\% less than the best container, Podman with a standard image (ratio 0.91, 90\% confidence interval [0.85, 0.97]).  
This difference is not significant after Holm correction, and the two environments are not equivalent within $\pm$10\%; however, a one-side non-inferiority test shows that the unikernel does not cost more than the container.
The families also interleave: Docker consumes 30\% more than Nanos on Firecracker ($p_H = 0.0026$), whereas Podman consumes 25\% less than Unikraft on Firecracker ($p_H < 0.001$).
The best members of the two families differ by 9\% (not significant), less than the differences within each family: Podman consumes 14\% less than Docker, and Nanos 22 to 32\% less than Unikraft (see Table~\ref{tab:contrasts}).
\emph{The choice within a family matters more than the choice of the family}, contrary to the usual framing of the containers-versus-unikernels question in the literature~\cite{goethalsFunctionalPerformanceBenchmark2022,moebiusAreUnikernelsReady2024,dinhtuanUnikernelsVsContainers2025}. 
The spread of points in Figure~\ref{fig:env-vs-bm} further shows that the best environment depends on the runtime it hosts; we return to this point in \S\ref{sec:rq3}.

\begin{figure}[htbp]
    \centering
    \includegraphics[width=\columnwidth]{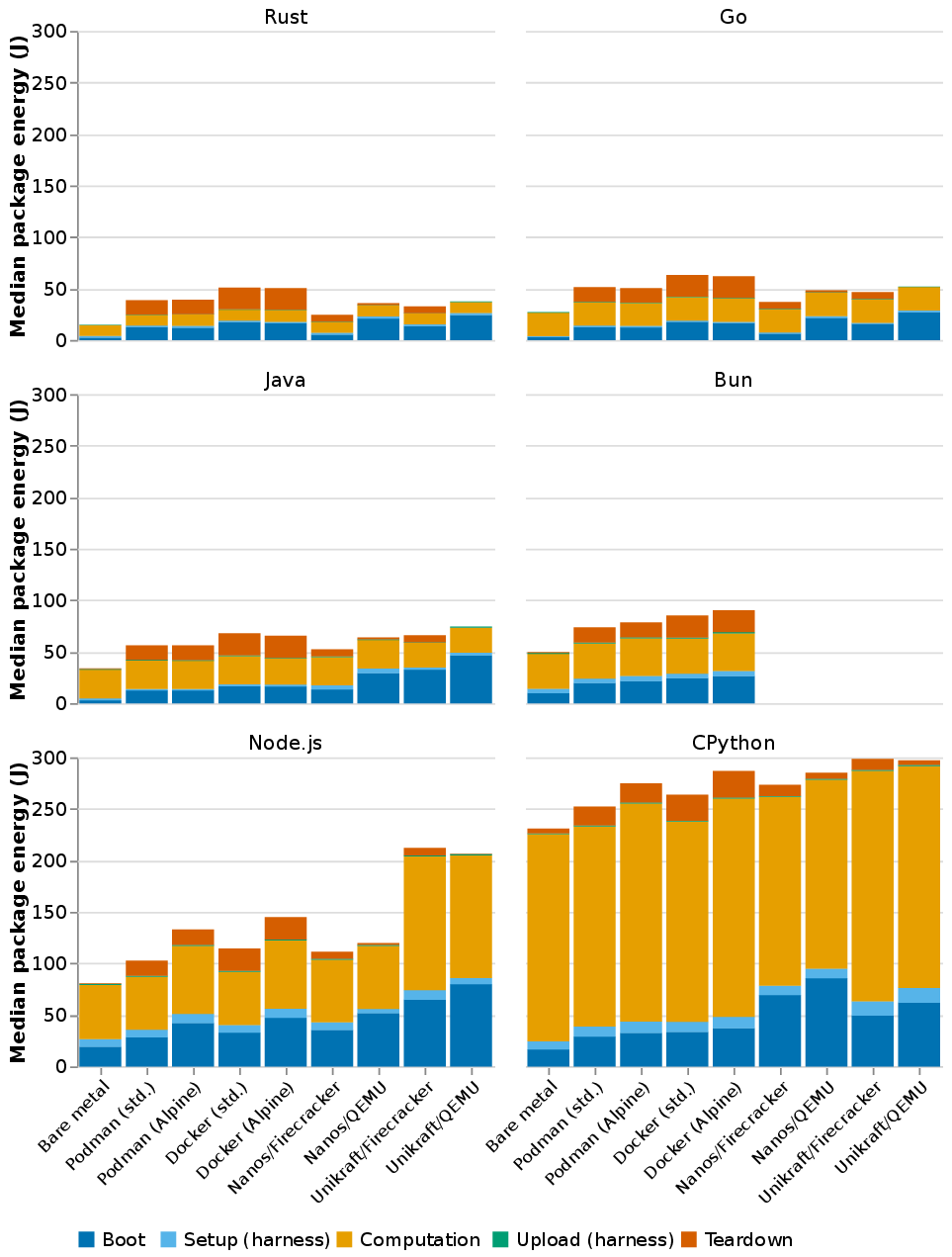}
    \caption{Median package energy per phase, for each language and environment (PyPy, deployable in only three environments, is omitted). Empty slots are configurations that cannot be deployed (Table~\ref{tab:runtime-matrix}). Computation dominates for CPython and Node.js, whereas boot and teardown dominate for Rust, Go, and Java; containers show a markedly higher teardown cost than unikernels and bare metal.}
    \label{fig:phase-breakdown}
\end{figure}

\subsubsection{Where environments differ: the lifecycle, including the teardown}\label{sec:lifecycle-bill}

\begin{table}[htbp]
  \caption{Targeted contrasts: one parameter changes at a time (geometric-mean ratios A/B over matched pairs, Wilcoxon signed-rank test with Holm correction). $^{***}p_H < 0.001$, $^{**}p_H < 0.01$, $^{*}p_H < 0.05$, n.s.\ not significant.}
  \label{tab:contrasts}
  \centering
  \scriptsize
  \begin{tabular}{lrrrrr}
    \toprule
    Contrast (A/B) & $n$ & $E_{cold}$ & Boot & Teardown & Compute \\
    \midrule
    Docker / Podman (glibc)        & 70 & 1.16$^{***}$ & 1.23$^{***}$ & 1.44$^{***}$ & 1.00 n.s. \\
    Docker / Podman (musl)         & 58 & 1.17$^{***}$ & 1.22$^{***}$ & 1.44$^{***}$ & 1.00 n.s. \\
    Alpine / standard (Docker)     & 58 & 1.06$^{*}$   & 1.07$^{**}$  & 1.00 n.s.    & 1.08$^{***}$ \\
    Alpine / standard (Podman)     & 58 & 1.07$^{***}$ & 1.10$^{***}$ & 1.00 n.s.    & 1.08$^{***}$ \\
    Firecr.\ / QEMU (Nanos)        & 50 & 0.82$^{***}$ & 0.46$^{***}$ & 3.84$^{***}$ & 0.98$^{***}$ \\
    Firecr.\ / QEMU (Unikraft)     & 50 & 0.95$^{**}$  & 0.69$^{***}$ & 29.1$^{***}$ & 1.02 n.s. \\
    Nanos / Unikraft (QEMU)        & 50 & 0.78$^{***}$ & 0.83$^{*}$   & 7.56$^{***}$ & 0.79$^{*}$ \\
    Nanos / Unikraft (Firecracker) & 50 & 0.68$^{***}$ & 0.55$^{*}$   & 1.00 n.s.    & 0.76$^{***}$ \\
    \bottomrule
  \end{tabular}
\end{table}

For the container engine and the hypervisor, environments barely affect the computation phase: in Table~\ref{tab:contrasts}, the compute column stays close to 1. 
The computation changes noticeably only when the software beneath the application differs. 
Only two contrasts change it, and in both, the code below the application differs: the C library (glibc vs.\ musl, $+8\%$ on the Alpine images) and the unikernel kernel (Nanos vs.\ Unikraft, 21 to 24\% less on Nanos). 
All remaining difference between environments therefore originate in the boot that precedes the computation and the teardown that follows it. 

\begin{figure}[htbp]
    \centering
    \includegraphics[width=\columnwidth]{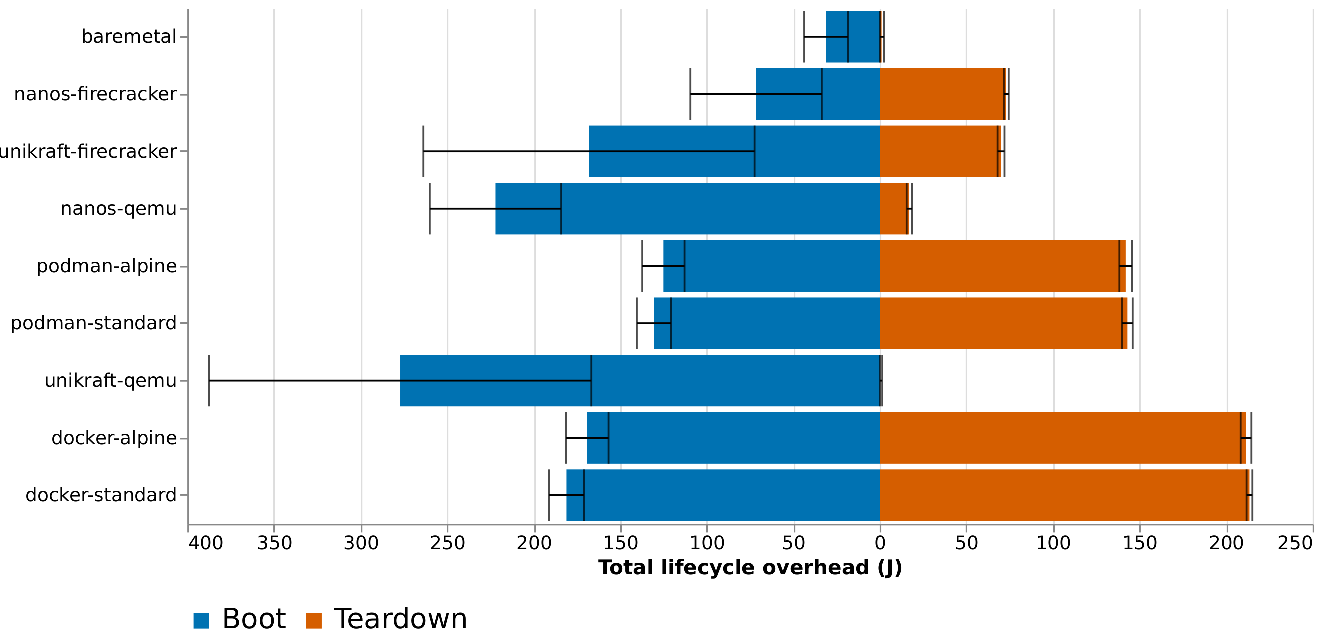}
    \caption{Median lifecycle energy of each environment for the \texttt{noop} workload, split into boot (left) and teardown (right). Environments are sorted by total cost. Teardown is negligible on bare-metal and QEMU-based unikernels, reaches about 6\,J under Firecracker, and matches or exceeds boot for containers (Docker, Podman).}
    \label{fig:lifecycle}
\end{figure}

Figure~\ref{fig:lifecycle} separates these two costs for the noop workload, which isolates the environment from any computation.
The teardown is the surprise: it represents a substantial share of the lifecycle. In containers, it amounts to 69 to 87\% of the boot energy (median over configurations), that is, 40 to 47\% of the lifecycle energy, and for Nanos on Firecracker it amounts to half of the boot energy. 
For the noop workload, it even exceeds the boot energy in containers.
It is negligible only under QEMU and on bare-metal.
Yet no cold-start study measures it, because the user never sees it. 

\textbf{Container engine: Docker pays 16\% more than Podman.} 
Over 70 matched pairs, Docker uses $16\%$ more cold-start energy than Podman ($p<10^{-11}$; Table~\ref{tab:contrasts}, first row). 
Its boot consumes $23\%$ more energy and its teardown $44\%$ more, while the compute phase is identical (ratio 1.00). 
The result reproduces on the musl images (58 pairs, ratio 1.17), indicating that it originates in the engine rather than in the image. 
Both engines rely on \texttt{runc}, so the difference reflects the cost of Docker's client-daemon architecture, most of which occurs during teardown and is therefore not captured by boot-only studies.

\textbf{Base image: musl slows the computation down.} 
The Alpine images boot slightly slower ($+6.8\%$ on Docker, $+9.7\%$ on Podman) and stop in the same amount of time (Table~\ref{tab:contrasts}, "Alpine / standard" rows). 
Their compute phase, however is $8\%$ slower than on the standard images, for both engines, with high significance ($p<10^{-3}$). 
A smaller image therefore provides no benefit in our setup, where the images already reside on the machine; its only visible effect is the C library, and that effect works \emph{against} Alpine.

\textbf{Hypervisor: Firecracker wins at boot, QEMU wins at teardown.} 
On the same unikernel, Firecracker boots much faster (Nanos: $\times2.2$, Unikraft: $\times1.5$) but destroys its VM much slower (Nanos: $\times3.8$; Unikraft: $\times29.1$; Table~\ref{tab:contrasts}, "Firecr. / QEMU" rows). 
The net effect on $E_{cold}$ still favors Firecracker ($-18\%$ on Nanos, $-5\%$ on Unikraft), and the computation is nearly identical on both.

This is the one case where optimizing latency and optimizing energy lead to different conclusions. For latency, only the boot matters, and Firecracker wins clearly on both unikernels. For energy, the teardown also counts: Firecracker still saves 18\% on Nanos, but only 5\% on Unikraft, several times less than the boot alone suggests. A platform designer who selects a hypervisor from boot benchmarks alone may therefore be optimizing the wrong objective.

\textbf{Unikernel: Nanos beats Unikraft.}
With the same hypervisor, Nanos uses $22\%$ less cold-start energy than Unikraft under QEMU, and $32\%$ less under Firecracker.
Surprisingly, this gain does not originate solely in the lifecycle: the computation itself is $21\%$ faster on Nanos under QEMU and $24\%$ faster under Firecracker (compute ratios 0.79 and 0.76, Table~\ref{tab:contrasts}).
The kernel beneath the application is therefore not neutral with respect to the computation.
Under Firecracker, Nanos also boots with less than half the energy of Unikraft (13.6\,J against 32.7\,J).
Under QEMU, Nanos's teardown is larger than Unikraft's (1.7\,J against 0.2\,J), but it remains negligible relative to its boot energy (29.5\,J).

\subsubsection{The longer the function, the less the environment counts}\label{sec:rq2-duration}
The last factor is not a technology but a property of the function itself. 
The lifecycle is a fixed cost paid per invocation (Table~\ref{tab:contrasts}), whereas computation energy grows with the amount of work performed, so the environment's share of $E_{cold}$ decreases as the function's duration increases. 
Our three input sizes span this range by design, from about 100\,ms to about 10\,s of computation. 
At 100\,ms, the lifecycle accounts for a median of 54\% of $E_{cold}$ under Docker with a standard image, and for up to 92\% (Rust on regex-redux). 
This median falls to 16\% at 1\,s, and to below 2\% in every environment at 10\,s; the largest share at 10\,s is 7.5\%, for a single configuration (Rust on k-nucleotide, Unikraft on QEMU).
\S\ref{sec:rq3} converts this observation into the duration beyond which the environment is amortized.

\begin{tcolorbox}[boxrule=0.4pt, breakable]
\textbf{Answer to RQ2.} Three factors, besides the language.
(i) The environment: its effect is real but narrow once bare metal is excluded (at most 56\% between the best and the worst virtualized environment), and it affects boot and teardown, not computation.
(ii) The layers inside it, often counterintuitive: Podman beats Docker ($-14\%$), the standard images beats Alpine ($-6$ to $-7\%$), Nanos beats Unikraft ($-22$ to $-32\%$), and Firecracker beats QEMU overall but not on teardown.
(iii) Function duration, which decides how much this matters: the lifecycle's share of $E_{cold}$ falls from a median of 54\% at 100\,ms (Docker with a standard image) to under 2\% at 10\,s.
Throughout, the underreported cost is the teardown: 40 to 47\% of the lifecycle energy in containers.
\end{tcolorbox}

\subsection{RQ3: What are the most energy-efficient configurations to adopt when using serverless functions?}\label{sec:rq3}
RQ1 and RQ2 study each factor separately. 
RQ3 puts them together and asks two questions. 
How long must a function run before the cost of its environment becomes negligible? 
And, is there a single configuration that is the most energy-efficient in every case?

\subsubsection{How long does it take to amortize the environment?}
Boot and teardown are fixed costs paid once per invocation, while computation energy grows with duration. 
To compare this fixed cost across configurations, we express it as a duration:
\begin{equation}
  \tau = \frac{E_{boot} + E_{teardown}}{E_{compute}} \times t_{compute}
  \label{eq:tau}
\end{equation}
$\tau$ is the computation time after which the computation has consumed as much energy as the lifecycle. 
In other words, starting and stopping the environment costs as much energy as $\tau$ of computation. 
For a function that computes for a duration $t$, the lifecycle represents a share $\tau / (t + \tau)$ of the energy: half of it when $t = \tau$, and less than 10\% when $t > 9\tau$. 
Unlike a ratio such as the vPUE~\cite{dasilvaAssessingPowerUsage2025}, $\tau$ does not favor slow code. 
A slow runtime has a long computation, which makes its lifecycle look small in proportion. 
$\tau$, instead, expresses the lifecycle energy as a duration of computation at the runtime's own power draw ($E_{compute}/t_{compute}$); as long as computation energy grows linearly with duration, this value is the same whether it is measured over 100\,ms or over 10\,s of computation.

\begin{figure}[htbp]
  \centering
  \includegraphics[width=\linewidth]{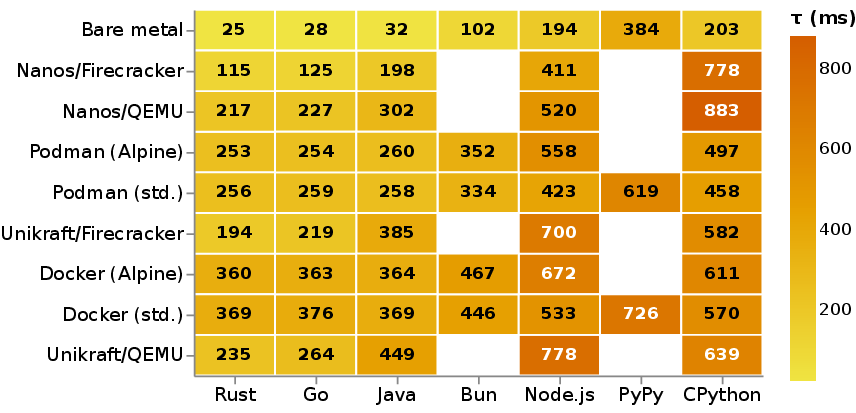}
  \caption{Time needed to amortize the environment, $\tau = \frac{E_{boot}+E_{teardown}}{E_{compute}} \times t_{compute}$ (median, 100\,ms). Lower is better; empty cells cannot be deployed (Table~\ref{tab:runtime-matrix}). The best virtualized environment depends on the runtime: Nanos on Firecracker for compiled code, Podman (std.\ image) for CPython, Bun, and PyPy.}
  \label{fig:tau}
\end{figure}

Figure~\ref{fig:tau} gives $\tau$ for every environment and language--runtime configuration. 
Three observations stand out. 
First, bare-metal amortizes almost immediately for compiled code: 25\,ms for Rust, 28\,ms for Go, and 32\,ms for Java. 
Second, every virtualized environment raises $\tau$ by about one order of magnitude: for Rust, from 115\,ms on Nanos/Firecracker to 369\,ms on Docker with a standard image. 
Third, $\tau$ also depends on the runtime. 
Interpreted and JIT runtimes start their interpreter during boot, so their $\tau$ is higher even on bare-metal (194\,ms for Node.js, 203\,ms for CPython, 384\,ms for PyPy), and it reaches 883\,ms for CPython on Nanos/QEMU.

These values are of the same order as the duration of real functions: about half of the functions in production run for less than one second on average~\cite{shahradServerlessWildCharacterizing2020}. 
For a function that computes for one second, the lifecycle represents 10\% of the energy with Rust on Nanos/Firecracker, but 27\% with Rust on Docker. 
To bring the lifecycle below 10\%, the same function must compute for at least 1\,s on Nanos/Firecracker, 3.3\,s on Docker, and 8\,s on Nanos/QEMU with CPython. 
For short functions, which are the common case, the environment is therefore not a detail: it can cost as much as the useful work.

\subsubsection{There is no one-size-fits-all configuration}
Rust on Nanos/Firecracker is one of the most energy-efficient configurations we measured when isolation is required: Rust has the lowest cold-start energy (mean rank 1.15, \S\ref{sec:rq1}), and Nanos/Firecracker is the best-ranked virtualized environment (\S\ref{sec:rq2}) with the lowest $\tau$ for Rust. 
However, neither choice holds in every case.

\emph{On the language axis}, Rust is first on average, but not on every workload (see Figure~\ref{fig:rank-stability}). 
Go takes first place on reverse-complement, and on pidigits CPython ranks 4th while Java falls to last, because the result depends on the big-integer library more than on the language. 
When a function relies on a native library, the best language is the one with the best library for that task.

\emph{On the environment axis}, bare-metal is always first, but it provides no isolation between functions. 
Among the environments that do isolate them, the best choice depends on the runtime (see Figure~\ref{fig:tau}). 
Nanos on Firecracker has the lowest $\tau$ for Rust (115\,ms), Go (125\,ms), Java (198\,ms), and, by a small margin, Node.js (411\,ms against 423\,ms on Podman). 
For CPython, the rank is reversed: Podman with a standard image is the best choice (458\,ms), while Nanos on Firecracker (778\,ms) and on QEMU (883\,ms) are the two worst. 
Bun and PyPy cannot run on unikernels, and Podman with a standard image is also the best choice for them (334\,ms and 619\,ms). 
The best virtualized environment for Rust is thus the second worst for CPython.

A configuration is therefore a pair, and its two parts must be chosen jointly. 
Rust on Nanos/Firecracker is a good answer for a team that can choose its language, but a team that must keep a Python code base should run it in a container, not in a unikernel.

\begin{tcolorbox}[boxrule=0.4pt, breakable]
\textbf{Answer to RQ3.}
No single configuration is best: Rust on Nanos/Firecracker is among the most efficient, but the best language depends on native libraries, and the best environment depends on the runtime (Nanos/Firecracker for compiled code, Podman with a standard image otherwise).
Choose the language-runtime pair first, then the environment.
The environment matters only while its lifecycle exceeds 10\% of the energy, that is, for functions that compute for less than $9\tau$: from about 1\,s (Rust on Nanos/Firecracker, $\tau = 115$\,ms) to 8\,s (CPython on Nanos/QEMU, $\tau = 883$\,ms).
Beyond that, changing the environment saves at most a few percent, whereas changing the language-runtime pair can save up to $11.4\times$.
\end{tcolorbox}

\section{Discussion}
\label{sec:discussion}

\subsection{Takeaways}
\label{sec:implications}

\textbf{The lifecycle cost grows with the number of invocations, not with the useful work.}
Each cold start pays the boot and teardown again, regardless of what the function computes. 
For a single invocation, this cost looks small: in median over our configurations, about 46\,J under Docker with a standard image, and 20\,J under Nanos on Firecracker. But a platform pays it at every cold start: over one million cold starts, this amounts to about 13\,kWh under Docker and 6\,kWh under Nanos on Firecracker, regardless of the work performed.
Short functions are the most affected, because their useful work is too small to hide this cost (\S\ref{sec:rq3}), and they are also the most common in production~\cite{shahradServerlessWildCharacterizing2020}. 
Reducing the lifecycle cost, or the number of cold starts, therefore matters even when the cost of one invocation seems negligible.

\textbf{Energy and latency do not always agree.}
In our setup, a sandbox uses a single core of a 12-core socket, so most package power is static, and a phase's energy is close to its duration multiplied by nearly constant power. 
As a consequence, most latency optimizations also reduce energy. 
The teardown is the exception: the user never waits for it, so latency studies ignore it, but the server still pays for it. 
This is why Firecracker, the clear winner on boot latency, loses most of its advantage on Unikraft once you count teardown (\S\ref{sec:lifecycle-bill}). An energy-aware platform must therefore measure the whole lifecycle, not only the part that the user sees.

\textbf{The teardown is also invisible in energy accounting.}
Serverless providers bill function execution time, and energy accounting approaches attribute energy to functions over the same window~\cite{sharmaAccountableCarbonFootprints2024, royHiddenCarbonFootprint2024}. 
The teardown happens after this window. 
Our measurements show that it can cost as much as the boot, so a share of the footprint of a function is currently attributed to no one. 
This is consistent with the observation of Basu Roy~\emph{et~al.}~\cite{royHiddenCarbonFootprint2024} that a large part of the footprint of a serverless platform never appears in the time billed to users.

\textbf{Runtime availability limits the benefit of unikernels.}
Unikernels offer hardware-level isolation at no energy premium for compiled code (\S\ref{sec:rq3}), but the most energy-efficient runtimes for JavaScript and Python (resp. Bun and PyPy) cannot be deployed on them, and CPython needs a longer time to amortize its lifecycle on a unikernel than in a container (see Figure~\ref{fig:tau}). 
For the languages that dominate serverless deployments, the obstacle to adopting unikernels is therefore not their energy cost, but the support of efficient runtimes.

\subsection{Threats to Validity}\label{sec:threats}
\textbf{Internal validity.} 
All artifacts are built from a single Nix flake with a pinned lock file, so two environments that share an implementation run the same binary. 
Not every combination could be measured: 114 of the 693 combinations (16.5\%) at 100\,ms could not be deployed, for the reasons given in \S\ref{sec:languages}.
Our statistical tests use matched blocks only, so these missing combinations reduce sample sizes without biasing the comparisons. 
Finally, supporting all nine environments required patching Unikraft (KVM paravirtual clock, \texttt{FUTEX\_REQUEUE}, epoll waking all waiters, \texttt{uk\_sys\_clock\_}\-\texttt{nanosleep}), so the Unikraft we measure is not the upstream version.

\textbf{Measurement artifacts.} 
Phase boundaries come from markers printed by the worker, and console buffering inside a VM can shift a marker. 
Two cases show this: the sleep window of noop on Unikraft/QEMU has a measured duration of zero, and the same workload on bare metal at 10\,s yields 69\,W, far outside the 85--98\,W range of every other measurement. 
In both cases, the energy is consistent across batches while the measured duration is not, which points to the marker rather than to a physical effect; we exclude both cases from the idle-power estimation.

\textbf{Construct validity.} 
We report the RAPL package and DRAM energy of the pinned socket; DRAM adds about 16\% and does not change any ranking. 
This excludes the second socket and the rest of the platform (fans, power supply), so our numbers describe what a function costs on a server that is already powered on, not its marginal cost on a busy node. 
$E_{cold}$ excludes the harness phases (connection to the queue and object store, job retrieval, and result upload), because they depend on external services rather than on the environment. 
Finally, the amortization time $\tau$ (\S\ref{sec:rq3}) assumes that the energy of the computation grows linearly with its duration. 
This holds for most configurations, but not all: the ratio of Node.js to Rust decreases with duration (Table~\ref{tab:lang}), suggesting that JIT warm-up makes the computation more efficient over time. 
For such runtimes, $\tau$ measured at 100\,ms is an upper bound.

\textbf{External validity.} 
All experiments use one node type (two Xeon Gold 6126), a single core, cold starts, and CPU- and memory-bound workloads whose only I/O goes through an object store. 
Two consequences matter. 
First, the high share of static power, which explains why energy follows duration so closely, is a property of this hardware generation. 
Second, providers do not run one sandbox at a time: with co-location, static power is shared across invocations and teardowns overlap other computations, so our per-invocation figures are upper bounds. 
Results may differ for I/O-heavy functions, for warm starts, or for other sandboxes (gVisor, Kata Containers, WebAssembly runtimes).

\textbf{Statistical validity.} 
Each configuration is summarized by its median over 60, 40, or 20 repetitions, depending on the input size. 
Noise is low (median coefficient of variation 6.4\%, median margin $\pm$1.6\%) and no distribution is bimodal, although 20 configurations exceed a coefficient of variation of 20\%, mostly noop on bare-metal, where the measured intervals are the shortest. 
We report the statistical power of every non-significant test, and we never conclude that two options are equivalent from a high $p$-value alone. 

\section{Conclusion and Future Work}\label{sec:conclusion}
Serverless platforms stack several software layers, whose cost in latency was much better known than their cost in energy. 
This paper measured both, phase by phase, across 9 execution environments, 7 language-runtime configurations, 11 workloads, and 3 input sizes. Three results stand out. 
First, the language-runtime pair has the largest effect: a factor of 11.4 between Rust and CPython, and changing only the runtime saves 31 to 48\%, as much as changing the whole execution environment. 
Second, apart from bare metal, environments differ almost entirely through their lifecycle. 
This fixed cost can exceed the energy of the computation itself for short functions, and up to half of it is teardown, a phase that cold-start studies ignore.
Third, there is no single best configuration: the best environment depends on the runtime it hosts. 


In the short term, we plan to extend our framework to warm starts and I/O-heavy workloads, and to exploit the idle baseline of noop for dynamic energy attribution. In the longer term, we aim to port Bun and PyPy to unikernels and to design an energy-aware scheduler that places each function on its best-suited sandbox.

\begin{acks}
This work received funding from the France 2030 program, managed by the French National Research Agency under grant agreement No. {\sf ANR-23-PECL-0003} (CARECloud) and {\sf ANR-26-CE25-5904-02} (EVERGREEN).  

This work is also done in the context of the Inria – Qarnot PULSE project: \url{https://www.inria.fr/en/pulse}, \url{https://defi-pulse.github.io/}.

Experiments presented in this paper were carried out using the Grid'5000 testbed, supported by a scientific interest group hosted by Inria and including CNRS, RENATER, several Universities, and other organizations (see \url{https://www.grid5000.fr}).
\end{acks}

\clearpage
\balance
\bibliographystyle{ACM-Reference-Format}
\bibliography{sample-base}
\end{document}